\documentclass{optica-article}

\journal{opticajournal} 

\articletype{Research Article}

\usepackage{subcaption}
\usepackage{braket}
\usepackage{siunitx}
\DeclareSIUnit{\hertz}{Hz}
\DeclareSIUnit{\sqrthz}{\sqrt{Hz}}
\DeclareSIUnit{\sens}{\tesla\per\sqrthz}
\usepackage{layouts}
\usepackage{todonotes}
\usepackage{booktabs}
\usepackage{float}
\floatstyle{plaintop}
\restylefloat{table}
\usepackage{booktabs}

\usepackage{layouts}
\usepackage{color}

\begin{document}

\title{High-contrast absorption magnetometry in the visible to near-infrared range with nitrogen-vacancy ensembles}


\author{Florian Schall\authormark{1,$\dagger$,*}, Felix A. Hahl\authormark{1,$\dagger$}, Lukas Lindner\authormark{1}, Xavier Vidal\authormark{1,2}, Tingpeng Luo\authormark{1}, Alexander M. Zaitsev\authormark{3}, Takeshi Ohshima\authormark{4,5}, Jan Jeske\authormark{1,*} and Rüdiger Quay\authormark{1}}

\address{
\authormark{1}Fraunhofer Institute for Applied Solid State Physics IAF, Tullastrasse 72, 79108 Freiburg, Germany\\
\authormark{2}TECNALIA, Basque Research and Technology Alliance (BRTA), 48160 Derio, Spain\\
\authormark{3}College of Staten Island (CUNY), 2800 Victory Blvd., Staten Island, NY 10312, USA\\
\authormark{4}National Institutes for Quantum Science and Technology (QST), 1233 Watanuki, Takasaki, Gunma 370-1292, Japan\\
\authormark{5}Department of Materials Science, Tohoku University, Aoba, Sendai, Miyagi, 980-8579, Japan}
\noindent
\authormark{$\dagger$}These authors contributed equally to this work.\\
\email{\authormark{*}florian.schall@iaf.fraunhofer.de, jan.jeske@iaf.fraunhofer.de}

\begin{abstract*} 
Magnetometry with nitrogen-vacancy (NV) centers has so far been measured via emission of light from NV centers or via absorption at the singlet transition at \SI{1042}{\nano\metre}. Here, we demonstrate a phenomenon of broadband optical absorption by the NV centers starting in the emission wavelength and reaching up to \SI{1000}{\nano\metre}. The measurements are enabled by a high-finesse cavity, which is used for room temperature continuous wave pump-probe experiments. The red to infrared probe beam shows the typical optically detected magnetic resonance (ODMR) signal of the NV spin with contrasts up to \SI{42}{\percent}. This broadband optical absorption is not yet reported in terms of NV magnetometry. We argue that the lower level of the absorbing transition could be the energetically lower NV singlet state, based on the increased optical absorption for a resonant microwave field and the spectral behavior. Investigations of the photon-shot-noise-limited sensitivity show improvements with increasing probe wavelength, reaching an optimum of \SI{7.5}{\pico\tesla\per\sqrt{\hertz}}. \\
The results show significantly improved ODMR contrast compared to emission-based magnetometry. This opens a new detection wavelength regime with coherent laser signal detection for high-sensitivity NV magnetometry. 
\end{abstract*}
\section{Introduction}
The negatively charged nitrogen-vacancy (NV) center is an optically active point defect in diamond \cite{doherty_paper, luo_paper}. It exhibits atomic-like behavior: The spin state of the electronic spin triplet state can be optically polarized and manipulated with a resonant microwave field. Spin-dependent intersystem crossing (ISC) and a spin-dependent fluorescence rate allow to read out the NV centers spin state, e.\,g. in an optically detected magnetic resonance (ODMR) experiment \cite{levine_microscope, schirhagel_paper}. These properties enable a wide range of applications in the field of quantum sensing. NV centers can be used to measure temperature \cite{acosta_temp}, pressure \cite{doherty_pressure}, stress \cite{kehayias_stress}, electric \cite{dolde_e_field} and magnetic fields \cite{schloss_vector}.\\
Especially in the field of magnetometry, the demand for highly sensitive and compact sensors has increased in recent years. Examples of magnetometry applications include the detection of magnetic anomalies \cite{keenan_anomaly}, nanoscale imaging of magnetic materials \cite{maletinsky_scanner, mathes_nanoscale} and the detection of neuronal signals in medicine (magnetoencephalography, MEG) \cite{boto_meg, zhang_meg, aslam_review}. NV-based magnetometers operate at room temperature under ambient magnetic fields, making them a promising alternative to other magnetometry technologies, which have high requirements on the experimental conditions \cite{aslam_review, degen_quantum_sensing}. NV magnetometers based on spin-dependent fluorescence detection currently can reach sensitivities in the range of $\SI{500}{fT\per\sqrt{\hertz}}$ \cite{barry_high_sens, wolf_paper, zhang_paper, graham_sens}. Due to small ODMR linewidths, most of these high-sensitivity sensors have only a small possible linear measurement regime of a few \SI{}{\micro\tesla}. However, NV magnetometers have a much larger range of background field strengths where they can be operated compared to other sensor technologies which often rely on zero-field environments as well as a large range of dynamic change of magnetic field strength the sensor can measure \cite{aslam_review,boto_meg}. The dynamic range can be quantified by the ODMR linewidth divided by the gyromagnetic ratio \cite{graham_sens} and can be as large as \SI{}{\milli\tesla} for NV center ensembles but is often traded to achieve improved sensitivity. \\
The concept of laser threshold magnetometry (LTM) predicts a further significant improvement to ODMR contrast and detection signal strength, resulting in better sensitivities combined with a high dynamic range \cite{jeske_ltm, dumeige_paper, jeske_stim, nair_amplification, nair_ltm, webb_ltm, hahl_paper, lindner_paper, bedford_paper}. This could enable further applications of NV magnetometry such as MEG. In LTM the NV centers are placed in an optical cavity and read out via cavity emission. They either provide magnetic-field-dependent gain \cite{jeske_ltm} or loss \cite{dumeige_paper}, resulting in a magnetic-field-dependent laser threshold. This concept predicts near-unity contrast with high-power and coherent laser output, resulting in an estimated sensitivity in the range of $\SI{1}{fT\per\sqrt{\hertz}}$ combined with a high dynamic range \cite{jeske_ltm}. A magnetic-field-dependent stimulated emission and an increased ODMR contrast of \SI{17}{\percent} were recently demonstrated with an external laser source at an emission wavelength of \SI{710}{\nano\metre} \cite{hahl_paper}. A photon-shot-noise-limited sensitivity of $\SI{30}{\pico\tesla\per\sqrt{\hertz}}$ with a dynamic range of \SI{250}{\micro\tesla} were measured \cite{hahl_paper}, demonstrating the potential of LTM experimentally for the first time. In a diode-assisted dual-media approach at \SI{690}{\nano\metre}, the first NV laser system and laser threshold created by optically pumping the NV diamond and the accessibility of the laser regime close to the threshold were shown \cite{lindner_paper}. Magnetic-field-dependent losses have also been demonstrated in continuous wave (cw) pump-probe experiments at the zero-phonon line (ZPL) of the infrared NV center singlet transition at a wavelength of \SI{1042}{\nano\metre} \cite{jensen_1042nm, chatzidrosos_1042nm, dumeige_1042nm}. Recently, a first experimental realization of infrared LTM at the singlet transition has been shown, demonstrating an improvement in sensitivity when approaching the laser threshold \cite{bedford_paper}. All these approaches are based on a high-finesse optical cavity, which significantly enhances the stimulated emission or absorption by the NV centers resulting in improved ODMR contrast. To optimize the signal-to-noise ratio the absorption-based approaches \cite{dumeige_paper, bedford_paper, jensen_1042nm, chatzidrosos_1042nm, dumeige_1042nm} drive the singlet transition resonantly due to maximized absorption at \SI{1042}{\nano\metre} \cite{kehayias_paper}. Pump-probe measurements in \cite{kehayias_paper, kehayias_supp} suggested a phonon-broadened absorption spectrum of the lower singlet state. This was used to perform an ODMR measurement with a probe wavelength at the first phonon sideband at \SI{912}{\nano\metre}. This firstly demonstrated a microwave-induced optical absorption besides the resonant wavelength of \SI{1042}{\nano\metre} of the singlet transition. However, the small signal-to-noise ratio required a lock-in amplifier in combination with a modulation of the pump laser. \\

\noindent In this work, we present cavity-enhanced cw pump-probe magnetometry with an NV ensemble pumped at \SI{532}{\nano\metre} and an external probe laser that is tunable from \SI{710}{\nano\metre} to \SI{1000}{\nano\metre}. Besides the stimulated emission at \SI{710}{\nano\metre} we find a broadband red to near-infrared absorption at room temperature, which shows a clear ODMR signal for all probe wavelengths. A continuous increase of ODMR contrast with increasing probe wavelength is observed. Contrast values of up to \SI{42}{\percent} in the entire curve and up to \SI{32}{\percent} for a single $m_S=0\rightarrow m_S=\pm 1$ spin transition are demonstrated. Coherent detection signals of a few \SI{}{\milli\watt} are possible, demonstrating the advantages of this cavity-based approach over typical fluorescence detection. The combination of high contrast and high detection signals enables an improved photon-shot-noise-limited sensitivity in the single-digit \unit{\pico\tesla\per\sqrt{\hertz}} range when increasing the probe wavelength. This sensitivity is limited by the available laser power of our probe laser and by increased optical NV-absorption, which occurs even in the absence of a microwave field when the NV centers are optically pumped.

\section{Experimental Setup and Methods}
\begin{figure}[htp]
	\centering
	\includegraphics[scale=1]{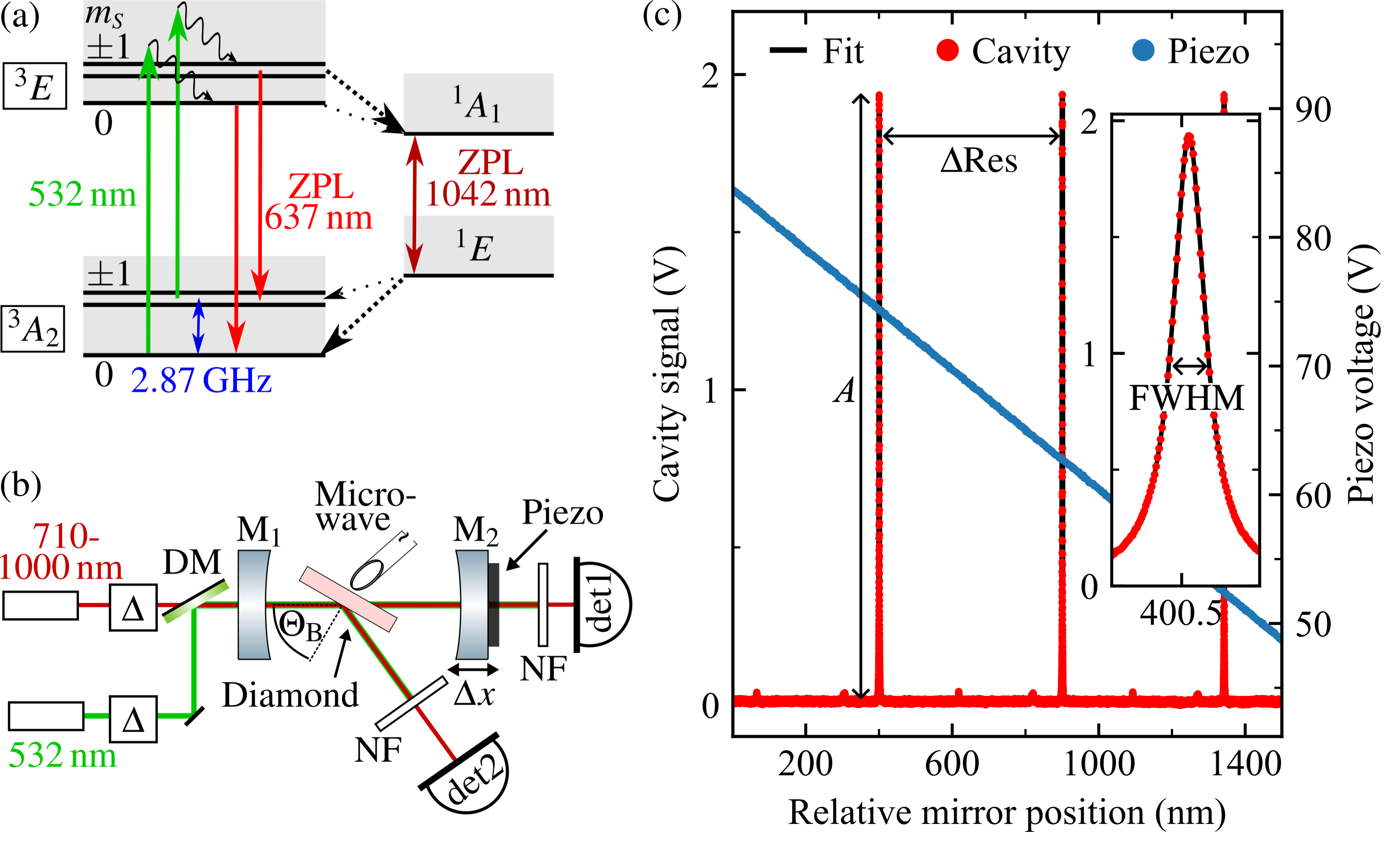}
	\caption{\textbf{(a)} Energy diagram of the negatively charged NV center in diamond. Optical transitions and the corresponding zero-phonon lines (ZPL) are indicated by solid arrows. The thickness of the dashed arrows indicates the strength of the intersystem crossing (ISC). The grey shading symbolizes the phonon broadening of the energy states. \textbf{(b)} Schematic of the experimental setup \cite{hahl_paper, hahl_phd}. The pump laser (green) and probe laser (red) are overlapped before the cavity with a dichroic mirror (DM). The $\Delta$ symbolizes the individual power, beam size and polarization adjustments for both lasers. The cavity signal is detected in transmission (det1) or in reflection (det2) by a power-calibrated variable gain photoreceiver with the pump laser being blocked by an optical filter (NF). \textbf{(c)} Typical cavity signal (red) on the oscilloscope as the voltage at the piezo is linearly varied (blue). The resonances are fitted with a Lorentzian function (black) defined by the amplitude $A$ and the resonance full width at half maximum (FWHM). The distance $\Delta\textrm{Res}$ between two resonances is given by $\Delta\textrm{Res}=\lambda_{\textrm{probe}}/2$. The probe wavelength is $\lambda_{\textrm{probe}}=\SI{1000}{\nano\metre}$ with an input power of $P_{\textrm{probe}}\approx\SI{425}{\milli\watt}$.} 
	\label{fig:levelstructure}
	\label{fig:setup}
	\label{fig:osci_example}
\end{figure}

\noindent The level structure of the negatively charged NV center is shown in Fig.\ref{fig:levelstructure}(a). The ground state $^3A_2$ is a spin triplet where the spin states are not degenerate due to spin-spin interaction. The NV center can be optically pumped to the excited triplet state $^3E$ with the green pump laser ($\lambda_{\textrm{pump}}=\SI{532}{\nano\metre}$). Optical pumping leads to spontaneous emission ranging from \SI{637}{\nano\metre} to \SI{800}{\nano\metre} \cite{levine_microscope}. The spin state can be prepared to $m_S=0$ utilizing spin-dependent ISC and readout by a spin-dependent fluorescence, as spin \num{1} exhibits a smaller fluorescence due to a stronger coupling to the less emitting singlet transition $^1A_1\rightarrow^1E$. This allows the detection of the magnetic-field-dependent resonance frequencies of a microwave field which drives the transition $m_S=0\rightarrow m_S=\pm 1$ via fluorescence detection for varying microwave frequencies (ODMR) \cite{levine_microscope}. Alternatively, the spin state can be detected by an absorption change of a probe beam driving the singlet transition $^1A_1\rightarrow^1E$ \cite{dumeige_1042nm}. A resonant microwave field results in an increased optical absorption due to a higher population of the absorbing energetically lower singlet state $^1E$. The optical absorption by this metastable state decreases drastically with an increasing temperature \cite{kehayias_supp, acosta_singlet}. This results in small and non-detectable signal changes at room temperature when only passing once through the diamond.

\noindent To improve the signal-to-noise ratio and to enhance the NV signal, a multi-pass approach with a high-finesse optical cavity is used \cite{hahl_paper,jensen_1042nm, chatzidrosos_1042nm}. The experimental setup is shown in Fig.\ref{fig:setup}(b). The diamond is placed in a linear optical cavity consisting of two plano-concave mirrors (ROC=\SI{30}{\milli\metre}) at a distance of $L\approx\SI{20}{\milli\metre}$. The mirrors are highly-reflectively coated ($R>$\SI{99.9}{\percent}) for the full bandwidth of the red probe laser and anti-reflectively coated ($R<$\SI{0.2}{\percent}) for the green pump laser. The cavity geometry results in a TEM$_\textrm{00}$ cavity mode with a beam waist of $\omega_{\textrm{cav}}\approx\SI{63}{\micro\metre}$ at the position of the diamond \cite{milloni_laser_physics}. The spot sizes and positions of both lasers are adjusted to achieve good overlap at the diamond position. The diamond is placed at Brewster's angle and both lasers are $p$-polarized to minimize the reflection losses at the diamond surface. The microwave antenna is a single loop antenna which requires high input power as it cannot be impedance-matched.  \\
\noindent The diamond is a (100) oriented commercial high-pressure-high-temperature (HPHT) diamond from ElementSix with a size of \num{3}x\num{3}x\SI{0.3}{\milli\metre\cubed}. Before the normal NV creation process the diamond was pretreated with a low-pressure-high-temperature (LPHT) annealing at \SI{1800}{\celsius} in an inert gas atmosphere (see \cite{hahl_paper,hahl_phd} for details). By this method, an ultra-low absorption of less than $\SI{0.01}{cm^{-1}}$ in the bandwidth of the probe laser is achieved, resulting in reduced losses in the cavity. This is an important point for improving the signal-to-noise ratio of the setup by increasing the cavity finesse. After the LPHT treatment the diamond was electron irradiated with a fluence of $\SI{1e18}{cm^{-1}}$ and an electron energy of \SI{2}{\mega\eV}. To create NV centers the diamond was again annealed for 2 hours at \SI{1000}{\celsius}. The NV concentration is approximately \SI{1.9}{ppm}, with only a small fraction populating the neutral charge state (<\SI{0.1}{ppm}). This high NV$^-$ concentration leads to a stronger NV signal, resulting in improved sensitivity \cite{schirhagel_paper, barry_paper}. \\

\noindent A cavity signal is generated by linearly moving the output cavity mirror (M$_2$) with a piezoelectric ring stack, thus varying the cavity length. A typical measurement signal is shown in Fig. \ref{fig:osci_example}(c). If the resonance criterion of the cavity is fulfilled, i.\,e. the cavity length is an integer multiple of the half probe wavelength \cite{milloni_laser_physics}, the cavity resonances are detected either in transmission (det1) or in reflection (det2) and visualized with an oscilloscope. The detectors are power-calibrated photoreceivers with variable gain, allowing a voltage-to-power conversion of the data. The Lorentzian-shaped resonances are defined by the amplitude $A$, and the full width at half maximum (FWHM), which is directly proportional to the losses in the cavity \cite{pollnau_resonators}. A dimensionless quantity to estimate the losses in the cavity is the finesse $\mathcal{F}$. It is defined as $\mathcal{F}=\Delta\textrm{Res}/\textrm{FWHM}$, where $\Delta\textrm{Res}=\lambda_{\textrm{probe}}/2$ is the distance between two resonances \cite{hahl_paper}. Utilizing high reflectivity mirrors, the low-absorbing diamond and the diamond orientation in Brewster's angle, a low-loss cavity setup with a high finesse of $\mathcal{F}>1200$ can be achieved (see \cite{hahl_paper, hahl_phd} for details). A high finesse leads to many round-trips of the probe laser through the cavity and thus through the diamond, which enhances the NV signal by a factor $N=2\mathcal{F}/\pi$ \cite{jensen_1042nm}. Such a multi-pass configuration significantly increases the signal-to-noise ratio, enabling the detection of small NV signals.

\section{Results and discussion}
\subsection{Broadband absorption magnetometry}
\begin{figure}[htp]
	\centering
	\includegraphics[scale=1]{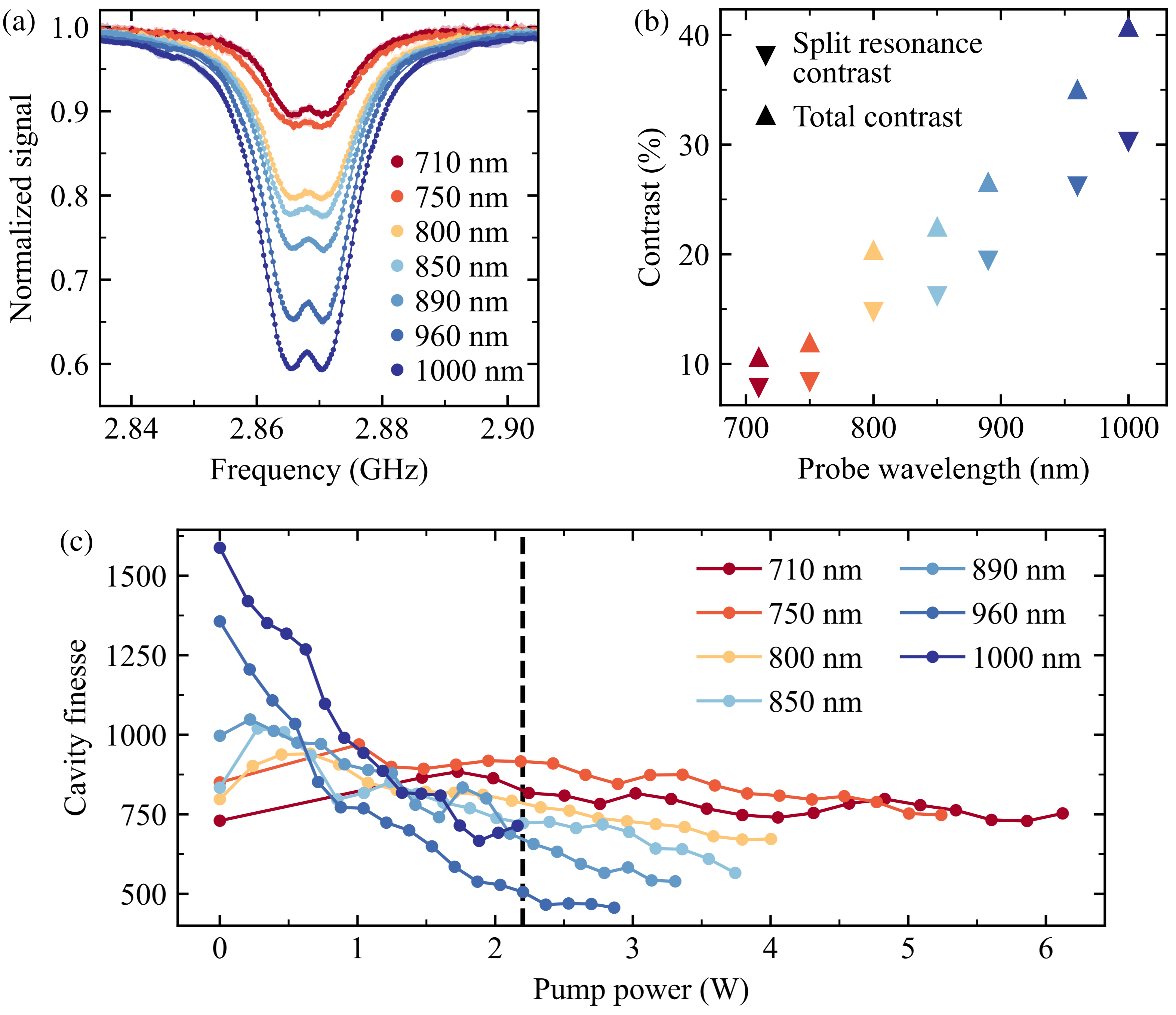}
	\caption{\textbf{(a)} Normalized ODMR spectra and \textbf{(b)} corresponding contrast for different probe wavelengths showing a strong increase in contrast for increasing wavelengths. The data points in (a) are the mean value with the standard deviation shown as shaded area which is small enough to mostly vanish behind the data points. The data is fitted with a double Lorentzian (solid lines). The pump and microwave powers used are comparable for the different probe wavelengths and are given by $P_{\mathrm{pump}}\approx\SI{2.2}{\watt}$ and $P_{\textrm{mw}}\approx\SI{32}{dBm}$. The probe power used is strongly dependent on the probe wavelength and varies between $P_{\mathrm{probe}}\approx\SI{0.25}{\watt}$ and $P_{\mathrm{probe}}\approx\SI{1}{\watt}$, but it is constant for each individual wavelength. \textbf{(c)} Cavity finesse for different probe wavelengths and variable pump powers showing the effect of pump-induced optical absorption, especially for high wavelengths. The data points are the rolling average of two neighboring measurements points. The measurement is noisy due to mechanical instabilities of the experimental setup. The dashed line symbolizes the pump power used for the ODMR spectra in (a). The pump power limits
	were chosen to investigate only the area of interest, thus minimizing the measurement time.} 
	\label{fig:odmr}
	\label{fig:contrast}  
	\label{fig:finesse} 
\end{figure}

As a first step, ODMR measurements were performed for different probe wavelengths to investigate the behavior of magnetic-field-dependent stimulated emission and possible absorption effects. The probe wavelengths were chosen to achieve near-equidistant sampling over the titanium-sapphire laser bandwidth and to achieve a stable wavelength locking of the probe laser. For each wavelength the cavity, diamond position and beam overlap were adjusted to realize the highest finesse, i.\,e. the lowest losses, resulting in the maximum cavity signal. The resulting ODMR spectra and the contrasts achieved for comparable pump and microwave powers are shown in Fig. \ref{fig:odmr}(a) and Fig \ref{fig:contrast}(b), respectively.\\
\noindent There is a strong increase in ODMR contrast with increasing probe wavelength. For a wavelength of \SI{1000}{\nano\metre} the contrast for a single spin transition, which we call split resonance, is $C>\SI{30}{\percent}$ and the total contrast is $C_{\mathrm{tot}}>\SI{40}{\percent}$. To the best of our knowledge, these high contrasts are a new record for an ensemble of NV centers. The high contrast was also achieved when applying a magnetic field along the (100) direction, to clearly split the $m_S=\num{0}\rightarrow m_S=\num{\pm 1}$ spin transitions (see Supplement S1 for details). 

\newpage\noindent For small probe wavelengths the high contrast can well be explained by magnetic-field-dependent stimulated emission of the NV centers into the cavity mode \cite{hahl_paper}. However, at high probe wavelengths the increased contrasts cannot entirely be explained by magnetic-field dependent stimulated emission due to vanishing spontaneous and thus stimulated emission. Instead, a broadband microwave-induced optical absorption must be the reason for the ODMR curves at high wavelengths. The lower state of the optical absorbing transition is populated by a resonant microwave field, since the optical absorption is increased (the transmission signal is decreased) for a microwave field at the NV spin resonance. Additionally, we know that a resonant microwave field increases the population in the $m_S = \pm 1$ triplet state and reduce the population in the $m_S=0$ triplet state. We also know that during cw optical pumping a resonant microwave field lead to an increased population in the singlet state $^1E$ due to the increased ISC from the $m_S=\pm 1$ state (see Fig. \ref{fig:levelstructure}(a)). Thus the lower state of the optical absorbing transition could be the energetically lower singlet state $^1E$. The upper state of the absorbing transition could be a phonon sideband of the upper singlet state or some other unknown state. The measurements in \cite{kehayias_paper} of the absorption spectrum of the lower singlet state $^1E$ using a lock-in amplifier support the assumption that the microwave-induced optical absorption comes from  driving the singlet transition off-resonantly. The increasing contrast with increasing probe wavelength is in good agreement with a stronger absorption by the lower singlet state $^1E$ for higher wavelengths \cite{kehayias_paper}. The weak singlet absorption at room temperature is strongly enhanced by the high-finesse cavity. A high finesse leads to many round-trips of the probe laser through the cavity and by that through the diamond, effectively increasing the absorption length, resulting in increased optical absorption \cite{jensen_1042nm}.\\
\noindent A measurement of this microwave-induced optical absorption at room temperature with such a high spectral bandwidth has not been realized before and is only possible because of the high-finesse cavity. The increase in cavity losses due to increased microwave-induced optical absorption is also shown by a drop of the cavity finesse for a resonant microwave field (see Supplement S2 for details). \\
\noindent To learn more about this effect, we determined the total signal loss in the cavity via measuring the cavity finesse for different NV pump powers and probe wavelengths. The results are shown in Fig. \ref{fig:finesse}(c). \\
Without pumping the NV centers the finesse increases for higher probe wavelengths, mainly because the intrinsic absorption of the diamond decreases at higher wavelengths, resulting in lower cavity losses and thus smaller FWHM (see \cite{hahl_phd} for details). Intrinsic absorption means probe laser absorption by, e.\,g. other defects in the diamond, without optically pumping the diamond with the green laser. For a pump power of $P_\mathrm{pump}\approx\SI{2.2}{\watt}$, which was used for the measurements in Fig. \ref{fig:odmr}(a) and is indicated with a dashed line, the finesse for a probe wavelength of \SI{1000}{\nano\metre} is $\mathcal{F}\approx\num{710}$. This leads to an enhancement of the optical absorption by $N\approx\num{450}$ because of an effective diamond length of $\approx\SI{13.5}{\centi\metre}$ due to $N$ passes through the diamond instead of \SI{0.3}{\milli\metre} for a single-pass. This increased absorption length results in the demonstrated high ODMR contrast due to microwave-induced optical absorption most likely by the lower singlet state. While the argument of increased contrast due to a high-finesse cavity applies to all wavelengths, we point out that at a pump power of \SI{2.2}{\watt} the finesse at \SI{1000}{\nano\metre} is actually lower than at \SI{750}{\nano\metre} (see Fig \ref{fig:finesse}(c)). An additional influence on the contrast is the variation of the absorption coefficient for different probe wavelengths. The high contrast at \SI{1000}{\nano\metre} implies a high absorption coefficient, which is in agreement with the absorption spectrum of the lower singlet state \cite{kehayias_paper}. The contrast is proportional to the product of the finesse and the absorption coefficient. For a small absorption coefficient the contrast depends linearly on the cavity finesse as it is experimentally demonstrated in Supplement S2 \cite{gagliardi_book}. \\
Nevertheless, optical losses are also induced by the pump laser, resulting in a decrease in finesse with increasing pump power. This pump-induced optical absorption broadens the FWHM of the cavity resonances due to increased losses, resulting in decreased cavity finesse (see Fig. \ref{fig:finesse}(c)). This loss channel could also be explained with a probe laser absorption by the lower singlet state $^1E$. With increasing pump power the lower singlet state becomes more populated, leading to more absorption losses and a decreased finesse. The pump-induced optical absorption shows a stronger influence for high probe wavelengths, resulting in a strong decrease of the finesse with increasing pump power. There is a strong increase in pump-induced optical absorption from \SI{710}{\nano\metre} to \SI{1000}{\nano\metre}. This is the same spectral behavior as that of the microwave-induced optical absorption, and it is in good agreement with the absorption spectrum of the lower singlet state \cite{kehayias_paper}. The pump-induced optical absorption severely limits the cavity finesse and thus the possible output power and the achievable ODMR contrast, due to a reduced number of round-trips in the cavity (see Supplement S2). However, a high cavity finesse can still be achieved with our setup, resulting in the very high ODMR contrast, which is promising for highly sensitive magnetometry. \\

\subsection{Sensitivity optimization}
\begin{figure}[htp]
	\centering
	\includegraphics[scale=1]{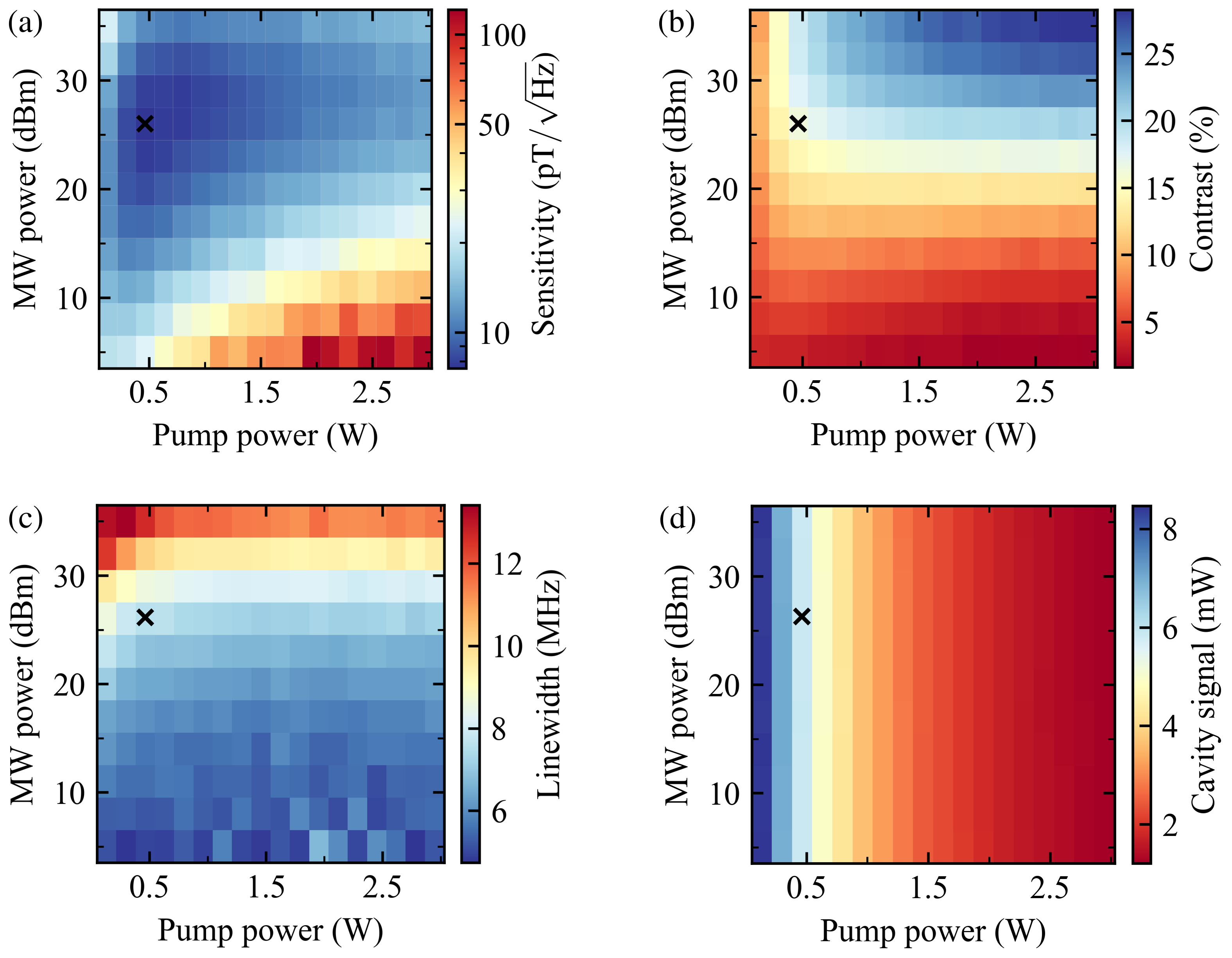}
	\caption{\textbf{(a)} Shot-noise-limited sensitivity, \textbf{(b)} split resonance contrast, \textbf{(c)} linewidth (FWHM) and \textbf{(d)} cavity signal of ODMR measurements for different pump and microwave (MW) powers showing a clear optimum in the sensitivity. The probe wavelength is \SI{960}{\nano\metre} with a power of $P_{\textrm{probe}}\approx\SI{830}{\milli\watt}$. The values are shown as a color plot, where red and blue symbolizes bad and good regarding sensitivity, respectively. The best sensitivity and the corresponding ODMR parameters are marked with a cross and are given in the article.}
	\label{fig:960nm_iteration}  
	\label{fig:960nm_sens}
	\label{fig:960nm_contrast}
	\label{fig:960nm_fwhm}
	\label{fig:960nm_baseline}
\end{figure}

In a second step, the photon-shot-noise-limited (PSNL) sensitivity $\eta_B$ in cw ODMR measurements was optimized to investigate the possible magnetic field sensitivity for the different probe wavelengths. The PSNL sensitivity for small contrasts is given by $\eta_B\propto\delta\nu/(C\sqrt{R})$ (see Supplement S3 for details) \cite{levine_microscope, hahl_paper}. Optimizing the sensitivity means minimizing the resonance linewidth $\delta\nu$ (FWHM) and maximizing the ODMR contrast $C$ and the detection rate $R=P_0/E_{\textrm{ph}}$. Here, $P_0$ is the cavity signal power (amplitude $A$ of the cavity resonance) and $E_{\textrm{ph}}$ is the photon energy of the detected photons \cite{hahl_paper}. For each probe wavelength, the pump and microwave powers were systematically iterated to find the optimum of the PSNL sensitivity. The limits for both powers were chosen to investigate only the area of interest, thus minimizing the measurement time. The results of the iteration for a probe wavelength of \SI{960}{\nano\metre} are shown in Fig. \ref{fig:960nm_iteration}.

\noindent One can clearly see an optimum for the PSNL sensitivity which is $\eta_{B}=\SI{7.5}{pT\per\sqrt{\hertz}}$ (see Fig. \ref{fig:960nm_sens}(a)). This sensitivity is achieved for a pump power of \SI{560}{\milli\watt} and a MW power of \SI{26}{dBm}. The corresponding ODMR curve shows a split resonance contrast of \SI{16}{\percent}, a linewidth of \SI{7.6}{\mega\hertz} and a cavity signal of \SI{5.8}{\milli\watt}. The sensitivity trends result from four counteracting processes, as can be seen from the results for the ODMR parameters. The split resonance contrast is maximized to more than \SI{28}{\percent} for high pump and microwave power (see Fig. \ref{fig:960nm_contrast}(b)). For high pump powers a good spin polarization of the absolute ground state with $m_S=0$ is achieved. Strong resonant microwave fields lead to high spin mixing between the spin states \num{0} and \num{1}. At high microwave powers the ODMR linewidth increases due to power broadening \cite{dreau_paper} (see Fig. \ref{fig:960nm_fwhm}(c)). This power broadening is much weaker for the pump laser because the pump intensity is far away from the saturation intensity ($s=I_\textrm{pump}/I_{\textrm{sat}}<\num{0.1}$) \cite{hahl_phd, dreau_paper}. A broad linewidth results in a worse sensitivity but a broader linear measurement regime, i.\,e. a high dynamic range of up to \SI{450}{\micro\tesla}. This value is estimated by the linewidth divided by the gyromagnetic ratio of the electron \cite{graham_sens}. Pump-induced optical absorption significantly reduces the cavity signal at high pump powers due to increased cavity losses (see Fig. \ref{fig:960nm_baseline}(d)). Detection signals of more than \SI{8}{\milli\watt} are achieved at low pump powers. These measurements with a high-finesse cavity demonstrate the advantages of high contrast and high-power detection signals, resulting in a good PSNL sensitivity combined with a high dynamic range. \\

\noindent Sensitivity optimization for other probe wavelengths show that for shorter wavelengths, the optimum PSNL sensitivity shifts to higher pump powers because the high contrast outweighs the weaker pump-induced optical absorption (see Supplement S3 for details). The optimum microwave power is almost independent of the probe wavelength. The influence of the probe power on the PSNL sensitivity was also investigated for the different probe wavelengths. For high probe powers, the ODMR contrast decreases up to a few percent, which can be explained by a saturation of the microwave-induced optical absorption (see Supplement S3 for details). However, for all wavelengths, the increase in cavity signal with increasing probe power ($P_0\propto P_{\textrm{probe}}$) outweighs the small decrease in contrast, resulting in the best PSNL sensitivity for the highest possible probe power. We note here that we were limited in probe power by our setup and higher powers could enable even better PSNL sensitivities (see Supplement S3 for details). \\

\begin{figure}[htp]
	\centering
	\includegraphics[scale=1]{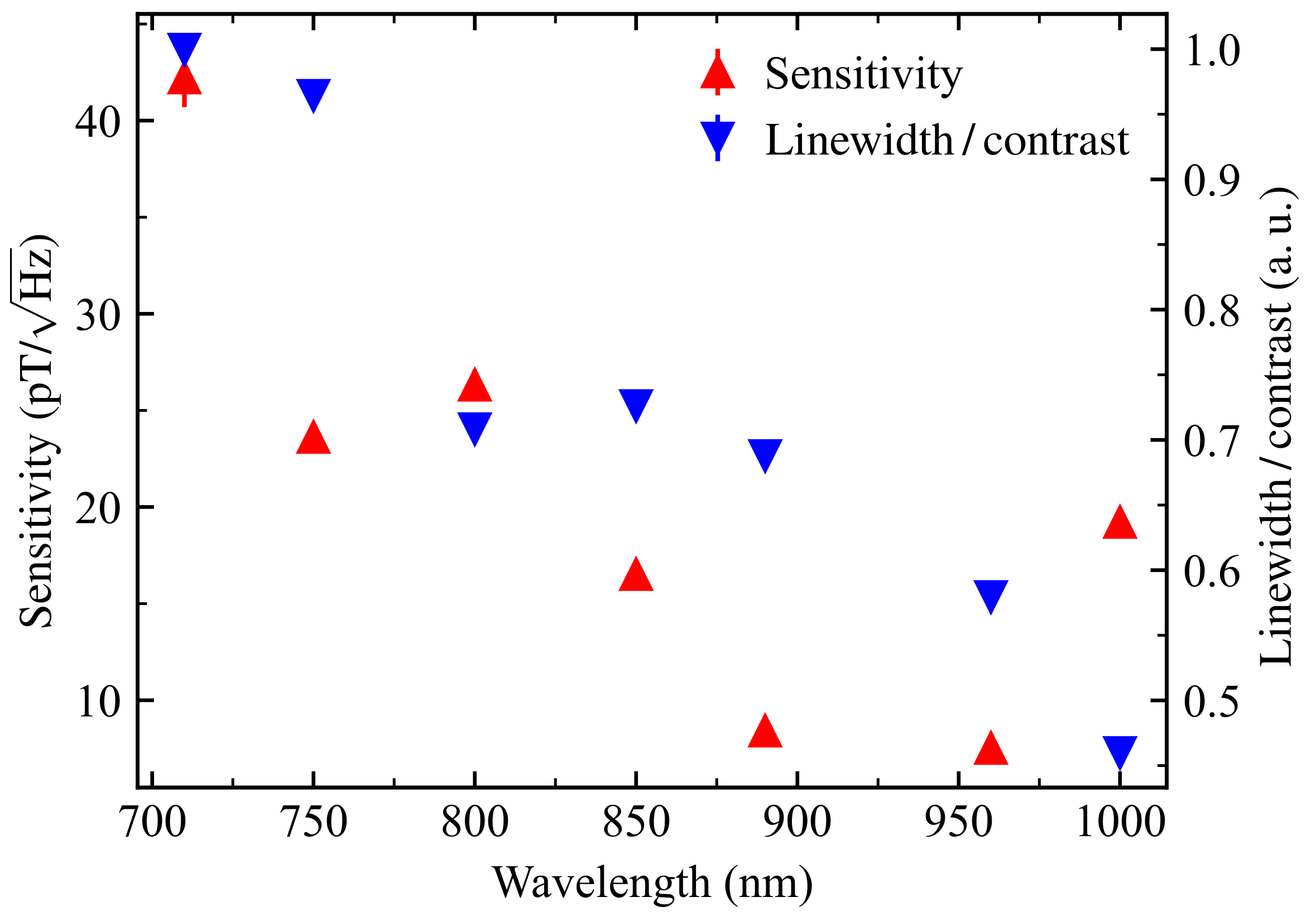}
	\caption{Optimized photon-shot-noise-limited sensitivity for different probe wavelengths (red). To neglect the influence of the probe power, the normalized linewidth over contrast ratio ($\delta\nu/C$) is also shown (blue). The data clearly shows that sensitivity is improved when going from the emission wavelength range to higher wavelengths, demonstrating the advantages of cavity-enhanced absorption magnetometry compared to emission magnetometry. The cavity signal was either detected in transmission or reflection depending on which signal was stronger. The laser and microwave powers used for the measurements and the ODMR parameters are given in Supplement S3.}
	\label{fig:comp_sens}
\end{figure}

\noindent The best achieved PSNL sensitivities for the investigated probe wavelengths are summarized in Fig. \ref{fig:comp_sens}. It can clearly be seen that increasing the probe wavelength improves the sensitivity. This is mainly due to the increased contrast at higher probe wavelengths. The sensitivity for a probe wavelength of \SI{1000}{\nano\metre} is worse because this wavelength is at the edge of the titanium-sapphire laser emission spectrum \cite{tisa}, resulting in a lower possible probe power of our laser. Neglecting the influence of probe power, i.\,e. looking at the ratio $\delta\nu/C$, there is a constant improvement with increasing probe wavelength (see blue points in Fig. \ref{fig:comp_sens}). The best sensitivity of $\eta_{B}=\SI{7.5}{pT\per\sqrt{\hertz}}$ in combination with a dynamic range of $\approx\SI{280}{\micro\tesla}$ was achieved with a probe wavelength of \SI{960}{\nano\metre}. All experimental and ODMR parameters for the different measurements are given in the Supplement S3. We note an improvement by a factor of three over the best achieved PSNL sensitivity at a probe wavelength of \SI{1042}{\nano\metre}, where a [111] bias field was applied and only one resonance was addressed \cite{chatzidrosos_1042nm}. The significant sensitivity boost with our setup is mainly achieved due to the higher cavity finesse, which increases the ODMR contrast to new record levels. The high finesse is made possible by the LPHT pretreatment of our diamond, which reduces the absorption coefficient in the wavelength range of the probe laser by more than an order of magnitude \cite{hahl_phd}. Compared to the best fluorescence-based sensors \cite{barry_high_sens, wolf_paper, zhang_paper}, we achieve a slightly worse sensitivity but our dynamic range is up to three orders of magnitudes higher allowing a wide range of possible applications. \\

\noindent  These measurements demonstrate a new and broad wavelength range for high sensitivity magnetometry combined with a high dynamic range. This opens a new field of possible laser sources and detectors for NV magnetometry. Especially for the experimental realization of LTM, our results demonstrate that approaches at higher red and infrared wavelengths promise to boost the sensitivity even further \cite{dumeige_paper, bedford_paper}. \\
  
\section{Conclusions}
We achieved strong ODMR signals from NV centers for wavelengths above the emission range, caused by broadband optical absorption. We show a continuous increase in ODMR contrast when the probe wavelength is increased from \SI{710}{\nano\metre} to \SI{1000}{\nano\metre}. Split resonance contrasts of more than \SI{32}{\percent} have been demonstrated with a total contrast of more than \SI{42}{\percent} in the absence of an external magnetic field. A reasonable assumption for the microwave-induced optical absorption is that we drive the NV singlet transition off-resonantly with our probe laser, resulting in increased optical absorption for a resonant microwave field due to a higher singlet population. This absorption by the lower singlet state is significantly enhanced by the high-finesse optical cavity, which results in many passes through the diamond, increasing the effective absorption length. \\
We also optimized the photon-shot-noise-limited sensitivity of the setup for the different probe wavelengths. For a probe wavelength of \SI{960}{\nano\metre}, we achieved the best photon-shot-noise-limit of $\eta_{B}=\SI{7.5}{pT\per\sqrt{\hertz}}$, combined with a high dynamic range of $\SI{280}{\micro\tesla}$. The sensitivity is limited by the achievable laser powers and pump-induced absorption that increases for higher probe wavelengths. This optical absorption is most likely also caused by the lower singlet state. The pump-induced optical absorption reduces the cavity finesse and thus the cavity signal and the achievable ODMR contrast due to microwave-induced optical absorption. \\ 
The measurements demonstrate an extended wavelength range for high-sensitivity quantum sensing with NV centers, which were not considered before. Compared to fluorescence-based setups, the presented results show a significantly improved contrast combined with a strong coherent detection signal. This is promising for further investigations to improve the sensitivity even more.\\

\noindent For the realization of a cavity-based magnetometer the sensitivity including technical noise is of interest. The current cavity was designed to have many degrees of freedom in alignment to investigate the different wavelength-dependent processes. Locking the optical cavity with a PID controller in combination with a compact and robust design will enable us to study and reduce technical noise sources. In addition, microwave delivery via Helmholtz coils or a microwave resonator can lead to a more homogeneous field over the large detection volume, resulting in narrowing the ODMR linewidth. The implementation of techniques, such as  double resonance magnetometry or hyperfine driving, could also boost the reachable sensitivity \cite{barry_paper, barry_high_sens}. Moving to higher probe wavelengths, e.\,g. to \SI{1042}{\nano\metre}, could on the one hand further improve the sensitivity due to stronger microwave-induced optical absorption by the lower singlet state \cite{kehayias_paper, kehayias_supp}. On the other hand, the presented cavity-approach enables a broadband spectroscopy of the singlet transition at room temperature, which was not possible before. \newline

\noindent
\textbf{Funding.} Bundesministerium für Bildung und Forschung (13N16485); Fraunhofer-Gesellschaft (QMag, QMag\,-\,Next level); Ministerium für Wirtschaft, Arbeit und Tourismus Baden-Württemberg (QMag).        

\noindent
\textbf{Disclosures.} FS, FAH, LL, TL, JJ: Fraunhofer-Gesellschaft zur Förderung der angewandten Forschung e.\,V. (P)

\noindent
\textbf{Data availability.} Data underlying the results presented in this paper are not publicly available at this time but may be obtained
from the authors upon reasonable requests.

\noindent
\textbf{Supplemental document} See Supplement for supporting content.

\bibliography{literature.bib}

\begin{thebibliography}{10}
\newcommand{\enquote}[1]{``#1''}

\bibitem{doherty_paper}
M.~W. Doherty, N.~B. Manson, P.~Delaney, \emph{et~al.}, \enquote{The
  nitrogen-vacancy colour centre in diamond,} {\protect\JournalTitle{Physics
  Reports}} \textbf{528} (2013).

\bibitem{luo_paper}
T.~Luo, L.~Lindner, J.~Langer, \emph{et~al.}, \enquote{Creation of
  nitrogen-vacancy centers in chemical vapor deposition diamond for sensing
  applications,} {\protect\JournalTitle{New Journal of Physics}} \textbf{24}
  (2022).

\bibitem{levine_microscope}
E.~V. Levine, M.~J. Turner, P.~Kehayias, \emph{et~al.}, \enquote{Principles and
  techniques of the quantum diamond microscope,}
  {\protect\JournalTitle{Nanophotonics}} \textbf{8} (2012).

\bibitem{schirhagel_paper}
R.~Schirhagel, K.~Chang, M.~Loretz, and C.~L. Degen, \enquote{Nitrogen-vacancy
  centers in diamond: Nanoscale sensors for physics and biology,}
  {\protect\JournalTitle{Annual Review of Physical Chemistry}} \textbf{65}
  (2013).

\bibitem{acosta_temp}
V.~M. Acosta, E.~Bauch, M.~P. Ledbetter, \emph{et~al.}, \enquote{Temperature
  dependence of the nitrogen-vacancy magnetic resonance in diamond,}
  {\protect\JournalTitle{Physical Review Letters}} \textbf{104} (2010).

\bibitem{doherty_pressure}
M.~W. Doherty, V.~V. Struzhkin, D.~A. Simpson, \emph{et~al.},
  \enquote{Electronic properties and metrology applications of the diamond
  $\textrm{NV}^{-}$ center under pressure,} {\protect\JournalTitle{Physical
  Review Letters}} \textbf{112} (2014).

\bibitem{kehayias_stress}
P.~Kehayias, M.~J. Turner, R.~Trubko, \emph{et~al.}, \enquote{Imaging crystal
  stress in diamond using ensembles of nitrogen-vacancy centers,}
  {\protect\JournalTitle{Phys. Rev. B}} \textbf{100} (2019).

\bibitem{dolde_e_field}
F.~Dolde, H.~Fedder, M.~W. Doherty, \emph{et~al.}, \enquote{Electric-field
  sensing using single diamond spins,} {\protect\JournalTitle{nature}}
  \textbf{7} (2011).

\bibitem{schloss_vector}
J.~M. Schloss, J.~F. Barry, M.~J. Turner, and R.~L. Walsworth,
  \enquote{Simultaneous broadband vector magnetometry using solid-state spins,}
  {\protect\JournalTitle{Physical Review Applied}} \textbf{10} (2018).

\bibitem{keenan_anomaly}
S.~T. Keenan, K.~R. Blay, and E.~J. Romans, \enquote{Mobile magnetic anomaly
  detection using a field-compensated high-t$_c$ single layer squid
  gradiometer,} {\protect\JournalTitle{Superconductor Science and Technology}}
  \textbf{24} (2011).

\bibitem{maletinsky_scanner}
P.~Maletinsky, S.~Hong, M.~S. Grinolds, \emph{et~al.}, \enquote{A robust
  scanning diamond sensor for nanoscale imaging with single nitrogen-vacancy
  centres,} {\protect\JournalTitle{nature nanotechnology}} \textbf{7} (2012).

\bibitem{mathes_nanoscale}
N.~Mathes, M.~Comas, R.~Bleul, \emph{et~al.}, \enquote{Nitrogen-vacancy center
  magnetic imaging of $\textrm{Fe}_3\textrm{O}_4$ nanoparticles inside the
  gastrointestinal tract of drosophila melanogaster,}
  {\protect\JournalTitle{Nanoscale Adv.}} \textbf{6} (2024).

\bibitem{boto_meg}
E.~Boto, N.~Holmes, J.~Leggett, \emph{et~al.}, \enquote{Moving
  magnetoencephalography towards real-world applications with a wearable
  system,} {\protect\JournalTitle{nature}} \textbf{555} (2018).

\bibitem{zhang_meg}
C.~Zhang, J.~Zhang, M.~Widmann, \emph{et~al.}, \enquote{Optimizing nv
  magnetometry for magnetoneurography and magnetomyography applications,}
  {\protect\JournalTitle{Frontiers in Neuroscience}} \textbf{16} (2023).

\bibitem{aslam_review}
N.~Aslam, H.~Zhou, E.~K. Urbach, \emph{et~al.}, \enquote{Quantum sensors for
  biomedical applications,} {\protect\JournalTitle{Nature Review Physics}}
  \textbf{5} (2023).

\bibitem{degen_quantum_sensing}
C.~L. Degen, F.~Reinhard, and P.~Cappellaro, \enquote{Quantum sensing,}
  {\protect\JournalTitle{Reviews of Modern Physics}} \textbf{89} (2017).

\bibitem{barry_high_sens}
J.~F. Barry, M.~H. Steinecker, S.~T. Alsid, \emph{et~al.}, \enquote{Sensitive
  ac and dc magnetometry with nitrogen-vacancy center ensembles in diamond,}
  {\protect\JournalTitle{Physical Review Applied}} \textbf{22} (2024).

\bibitem{wolf_paper}
T.~Wolf, P.~Neumann, K.~Nakamura, \emph{et~al.}, \enquote{Subpicotesla diamond
  magnetometry,} {\protect\JournalTitle{Physical Review Letters}} \textbf{5}
  (2015).

\bibitem{zhang_paper}
C.~Zhang, F.~Shagieva, M.~Wildmann, \emph{et~al.}, \enquote{Diamond
  magnetometry and gradiometry towards subpicotesla dc field measurement,}
  {\protect\JournalTitle{Physical Review Applied}} \textbf{14} (2021).

\bibitem{graham_sens}
S.~M. Graham, A.~T. M.~A. Rahman, L., \emph{et~al.}, \enquote{Fiber-coupled
  diamond magnetometry with an unshielded sensitivity of 30
  pt/$\sqrt{\textrm{hz}}$,} {\protect\JournalTitle{Physical Review Applied}}
  \textbf{19} (2023).

\bibitem{jeske_ltm}
J.~Jeske, J.~H. Cole, and A.~D. Greentree, \enquote{Laser threshold
  magnetometry,} {\protect\JournalTitle{New Journal of Physics}} \textbf{18}
  (2015).

\bibitem{dumeige_paper}
Y.~Dumeige, J.~F. Roch, F.~Bretenaker, \emph{et~al.}, \enquote{Infrared laser
  threshold magnetometry with a nv doped diamond intracavity etalon,}
  {\protect\JournalTitle{Optics Express}} \textbf{24} (2019).

\bibitem{jeske_stim}
J.~Jeske, D.~W.~M. Lau, X.~Vidal, \emph{et~al.}, \enquote{Stimulated emission
  from nitrogen-vacancy centres in diamond,} {\protect\JournalTitle{nature
  communications}} \textbf{8} (2017).

\bibitem{nair_amplification}
S.~R. Nair, L.~J. Rogers, X.~Vidal, \emph{et~al.}, \enquote{Amplification by
  stimulated emission of nitrogen-vacancy centres in a diamond-loaded fibre
  cavity,} {\protect\JournalTitle{Nanophotonics}} \textbf{9} (2020).

\bibitem{nair_ltm}
S.~R. Nair, L.~J. Rogers, D.~J. Spence, \emph{et~al.}, \enquote{Absorptive
  laser threshold magnetometry: combining visible diamond raman lasers and
  nitrogen-vacancy centres,} {\protect\JournalTitle{Materials for Quantum
  Technologies}} \textbf{1} (2021).

\bibitem{webb_ltm}
J.~L. Webb, A.~F.~L. Poulsen, R.~Staacke, \emph{et~al.}, \enquote{Laser
  threshold magnetometry using green-light absorption by diamond nitrogen
  vacancies in an external cavity laser,} {\protect\JournalTitle{Physical
  Review A}} \textbf{103} (2021).

\bibitem{hahl_paper}
F.~A. Hahl, L.~Lindner, X.~Vidal, \emph{et~al.},
  \enquote{Magnetic-field-dependant stimulated emission from nitrogen-vacancy
  centers in diamond,} {\protect\JournalTitle{Science Advances}} \textbf{8}
  (2022).

\bibitem{lindner_paper}
L.~Lindner, F.~A. Hahl, L.~Tingpeng, \emph{et~al.}, \enquote{Dual-media laser
  system: Nitrogen vacancy diamond and red semiconductor laser,}
  {\protect\JournalTitle{ScienceAdvances}} \textbf{10} (2024).

\bibitem{bedford_paper}
N.~S. Gottesman, M.~A. Slocum, G.~A. Sevison, \emph{et~al.}, \enquote{Infrared
  vertical external cavity surface emitting laser threshold magnetometer,}
  {\protect\JournalTitle{Applied Physics Letters}} \textbf{124} (2024).

\bibitem{jensen_1042nm}
K.~Jensen, N.~Leefer, A.~Jarmola, \emph{et~al.}, \enquote{Cavity-enhanced
  room-temperature magnetometry using absorption by nitrogen-vacancy centers in
  diamond,} {\protect\JournalTitle{Physical Review Letters}} \textbf{112}
  (2014).

\bibitem{chatzidrosos_1042nm}
G.~Chatzidrosos, A.~Wickenbrock, L.~Bougas, \emph{et~al.}, \enquote{Miniature
  cavity-enhanced diamond magnetometer,} {\protect\JournalTitle{Physical Review
  Applied}} \textbf{8} (2017).

\bibitem{dumeige_1042nm}
Y.~Dumeige, M.~Chipaux, V.~Jacques, \emph{et~al.}, \enquote{Magnetometry with
  nitrogen-vacancy ensembles in diamond based on infrared absorption in a
  doubly resonant optical cavity,} {\protect\JournalTitle{Physical Review B}}
  \textbf{87} (2013).

\bibitem{kehayias_paper}
P.~Kehayias, M.~W. Doherty, D.~English, \emph{et~al.}, \enquote{Infrared
  absorption band and vibronic structure of the nitrogen-vacancy center in
  diamond,} {\protect\JournalTitle{Physical Review B}} \textbf{88} (2013).

\bibitem{kehayias_supp}
P.~Kehayias, M.~W. Doherty, D.~English, \emph{et~al.}, \enquote{Supplemenatry
  materials to: Infrared absorption band and vibronic structure of the
  nitrogen-vacancy center in diamond,} {\protect\JournalTitle{Physical Review
  B}} \textbf{88} (2013).

\bibitem{acosta_singlet}
V.~M. Acosta, A.~Jarmola, E.~Bauch, and D.~Budker, \enquote{Optical properties
  of the nitrogen-vacancy singlet levels in diamond,}
  {\protect\JournalTitle{Physical Review B}} \textbf{82} (2010).

\bibitem{hahl_phd}
{Felix Anton Hahl}, \enquote{Cavity-enhanced magnetic-field sensing via
  stimulated emission from nitrogen-vacancy centres in diamond,} Ph.D. thesis,
  Albert-Ludwigs-Universität Freiburg im Breisgau (2022).

\bibitem{milloni_laser_physics}
P.~W. Milonni and J.~H. Eberly, \emph{Laser Physics} (John Wiley and Sons,
  2010).

\bibitem{barry_paper}
J.~F. Barry, J.~M. Schloss, E.~Bauch, \emph{et~al.}, \enquote{Sensitivity
  optimization for nv-diamond magnetometry,} {\protect\JournalTitle{Reviews of
  modern Physics}} \textbf{92} (2020).

\bibitem{pollnau_resonators}
M.~Pollnau and M.~Eichhorn, \enquote{Spectral coherence, part i:
  Passive-resonator linewidth, fundamental laser linewidth and schawlow-townes
  approximation,} {\protect\JournalTitle{Progress in Quantum Electronics}}
  \textbf{72} (2020).

\bibitem{gagliardi_book}
G.~Gagliardi and H.-P. Loock, \emph{Cavity-Enhanced Spectroscopy and Sensing}
  (Springer Berlin, 2016).

\bibitem{dreau_paper}
A.~Dreau, M.~Lesik, L.~Rondin, \emph{et~al.}, \enquote{Avoiding power
  broadening in optically detected magnetic resonance of single nv defects for
  enhanced dc magnetic field sensitivity,} {\protect\JournalTitle{Physical
  Review B}} \textbf{84} (2011).

\bibitem{tisa}
H.~Burton, C.~Debardelaben, W.~Amir, and A.~Planchon, \enquote{Temperature
  dependence of ti:sapphire fluorescence spectra for the design of cryogenic
  cooled ti:sapphire cpa laser,} {\protect\JournalTitle{Optics Express}}
  \textbf{25} (2017).

\end{thebibliography}

\end{document}